\documentclass[prl,twocolumn,letterpaper,superscriptaddress]{revtex4-1}
\usepackage[space]{grffile}
\usepackage{bm,graphicx,graphics,amsmath,amssymb,bm,epsfig,color}
\usepackage{euscript,tabularx}
\usepackage{longtable}
\usepackage{float}
\usepackage[draft]{todonotes}

\graphicspath{{./figs/final/}}

\begin{document}



\title{Electrical properties of single crystal Yttrium Iron Garnet
  ultra-thin films at high temperatures}

\author{N. Thiery}
\affiliation{SPINTEC, CEA-Grenoble, CNRS and Universit\'e Grenoble Alpes, 38054
  Grenoble, France}

\author{V. V. Naletov} 
\affiliation{SPINTEC, CEA-Grenoble, CNRS and Universit\'e Grenoble Alpes, 38054
  Grenoble, France}
\affiliation{Institute of Physics, Kazan Federal University, Kazan
    420008, Russian Federation}

\author{L. Vila}
\affiliation{SPINTEC, CEA-Grenoble, CNRS and Universit\'e Grenoble Alpes, 38054
  Grenoble, France}

\author{A. Marty}
\affiliation{SPINTEC, CEA-Grenoble, CNRS and Universit\'e Grenoble Alpes, 38054
  Grenoble, France}

\author{A. Brenac}
\affiliation{SPINTEC, CEA-Grenoble, CNRS and Universit\'e Grenoble Alpes, 38054
  Grenoble, France}

\author{J.-F. Jacquot}
\affiliation{SPINTEC, CEA-Grenoble, CNRS and Universit\'e Grenoble Alpes, 38054
  Grenoble, France}

\author{G. de Loubens} 
\affiliation{SPEC, CEA-Saclay, CNRS, Universit\'e Paris-Saclay,
  91191 Gif-sur-Yvette, France}

\author{M. Viret}
\affiliation{SPEC, CEA-Saclay, CNRS, Universit\'e Paris-Saclay,
  91191 Gif-sur-Yvette, France}

\author{A. Anane} 
\affiliation{Unit\'e Mixte de Physique CNRS, Thales, Universit\'e
  Paris-Saclay, 91767 Palaiseau, France}

\author{V. Cros}
\affiliation{Unit\'e Mixte de Physique CNRS, Thales, Universit\'e
  Paris-Saclay, 91767 Palaiseau, France}

\author{J. Ben Youssef} 
\affiliation{LabSTICC, CNRS, Universit\'e de Bretagne Occidentale,
  29238 Brest, France}

\author{V. E. Demidov}
\affiliation{Department of Physics, University of Muenster, 48149 Muenster, Germany}

\author{S. O. Demokritov} 
\affiliation{Department of Physics, University of Muenster, 48149 Muenster, Germany}
\affiliation{Institute of Metal Physics, Ural Division of RAS,
  Yekaterinburg 620041, Russian Federation}

\author{O. Klein}
\email[Corresponding author:]{ oklein@cea.fr}
\affiliation{SPINTEC, CEA-Grenoble, CNRS and Universit\'e Grenoble Alpes, 38054
  Grenoble, France}

\date{\today}

\begin{abstract}
  We report a study on the electrical properties of 19~nm thick
  Yttrium Iron Garnet (YIG) films grown by liquid phase epitaxy. The
  electrical conductivity and Hall coefficient are measured in the
  high temperature range [300,400]~K using a Van der Pauw four-point
  probe technique. We find that the electrical resistivity decreases
  exponentially with increasing temperature following an activated
  behavior corresponding to a band-gap of $E_g\approx 2$~eV,
  indicating that epitaxial YIG ultra-thin films behave as large gap
  semiconductor, and not as electrical insulator. The resistivity
  drops to about $5\times 10^3$~$\Omega \cdot \text{cm}$ at
  $T=400$~K. We also infer the Hall mobility, which is found to be
  positive ($p$-type) at 5~cm$^2$/(V$\cdot$sec) and about independent
  of temperature. We discuss the consequence for non-local transport
  experiments performed on YIG at room temperature. These electrical
  properties are responsible for an offset voltage (independent of the
  in-plane field direction) whose amplitude, odd in current, grows
  exponentially with current due to Joule heating. These electrical
  properties also induce a sensitivity to the perpendicular component
  of the magnetic field through the Hall effect. In our lateral
  device, a thermoelectric offset voltage is produced by a temperature
  gradient along the wire direction proportional to the perpendicular
  component of the magnetic field (Righi-Leduc effects).
\end{abstract}

\maketitle

\newcommand\encircle[1]{\tikz[baseline=(char.base)]{
            \node[shape=circle,draw,inner sep=-0.2pt] (char) {#1};}}

The recent discovery that spin orbit effects \cite{valenzuela06,
  kajiwara10, miron11, rojas13a} could allow to generate or to detect
pure spin currents circulating in an adjacent magnetic layer has
triggered a renewed interest for magnon transport in magnetic oxides,
and in particular Yttrium Iron garnet, Y$_3$Fe$_5$O$_{12}$ (YIG)
\cite{kajiwara10, wang11, padron-hernandez11, chumak12, hahn13,
  kelly13, hamadeh14b, collet16, lauer16, Du2017, Wesenberg2017,
  Thiery2017}, the material with the lowest known magnetic damping in
nature. It confers to YIG the unique ability to propagate the spin
information on the largest possible distance. Moreover, as YIG is an
electrical insulator, all spurious effects associated with electrical
transport properties are absent, which simplifies greatly the
interpretation of the measurements.

The latest studies on the magnon transport properties of YIG
concentrate on the strong out-of-equilibrium regime where large spin
currents are induced in the YIG either by way of spin transfer torque
\cite{Wesenberg2017, Thiery2017} or by temperature gradients
\cite{Safranski2017,Lauer2016}. When performed at room temperature,
this involves heating the YIG material well above 300~K. One possible
concern is the potential increase of its electrical conductivity at
high temperature. Indeed, it has been known since the seventies
\cite{Larsen1976, Metselaar1978, Lal1982, Petrov1986, Sirdeshmukh1998,
  Modi2014} that the electrical resistivity of doped YIG could
decrease by several orders of magnitude at high temperature due to the
presence of impurities. In the case of ultra-thin films defects could
come from the growth method or from the 2 interfaces and potentially
lead to a spurious charge conduction channel when heated well above
300~K. In order to clarify this point, we propose to investigate the
evolution of the electrical properties of single crystal Yttrium Iron
Garnet ultra-thin films at high temperatures.

Before describing the experimental procedure, we would like to recall
that YIG is a ferrimagnet, which has an uncompensated magnetic moment
on the Fe$^{3+}$ ions, found on octahedral and tetrahedral coordinate
sites, both coupled by super-exchange. Studies on Ca and Si doped YIG
\cite{Larsen1976} have established that Fe$^{2+}$ and Fe$^{4+}$ ions
are formed if tetravalent or respectively divalent impurities are
added to the YIG, which could then lead to an electrical conduction
via the charge transfer mechanism, respectively $p$-type and
$n$-type. In that case, the doped YIG behaves as a large gap
semiconductor with a charge conductivity following an activation
mechanism. At the present stage, different studies disagree about the
microscopic mechanism at play for the electronic conduction inside
doped YIG, whether it follows a localized hopping model, through a
small polaron conduction \cite{Sirdeshmukh1998} or rather a band
model, through a large polaron conduction \cite{Larsen1976}. It is
also known that the value of the magnetic damping coefficient of YIG
is very sensitive to the doping level. This is because the charge
transfer between the mixed valence iron ions is associated to a potent
magnetic relaxation process, known as the valence exchange relaxation
\cite{Sparks1969}. So far this mechanism activated by impurities,
appears in the form of a large enhancement of the magnetic damping,
usually around liquid nitrogen temperature, where the fluctuation rate
of the charge transfer matches the Larmor frequency. This effect is
usually minimized by growing YIG crystals from ultra-pure
materials. Quite remarkably YIG can usually be synthesized in large
volume in the form of a single crystal with almost no atomic disorder.
It has been reported that the resistivity of bulk ultra-pure YIG can
be as large as $10^{12}$~$\Omega \cdot \text{cm}$ at room temperature
\cite{Metselaar1978}.

But, as explained in the introduction, recent interest on spin
transfer effects in YIG have required an effort to develop high
quality YIG material in the form of ultra-thin (below 20~nm) films
(thickness should be compared here relatively to the YIG unit cell,
which is 1.238~nm). This is because spin transfer effect is an
interfacial phenomenon and consequently its efficiency increases with
decreasing thickness of the magnetic layer. Three growth techniques
have so far allowed to produce good quality ultra-thin YIG films:
sputtering \cite{fang2017, Kang2005, wang13}; pulsed laser deposition
\cite{kelly13,sun12,onbasli14,Hauser2016}; and liquid phase epitaxy
(LPE) \cite{Thiery2017, Dubs2017, hahn13}. These films are usually
grown on Gadolinium Gallium Garnet, Gd$_3$Ga$_5$O$_{12}$ (GGG)
substrates, which provides the necessary lattice matching to achieve
epitaxial growth. For all these three growth processes, the quality of
the YIG films deteriorates as the film thickness decreases
\cite{kelly13, Dubs2017}. This deterioration is an inherent
consequence of an increasing surface to volume ratio, which
substantially enhances the possibilities for defects and impurities to
be introduced into the YIG, through the two surfaces (contamination,
intermixion of the species at the surface or unrelaxed strains in the
film thickness), which leads to lower spontaneous magnetization and an
out-of-plane anisotropy accompanied or not by an increase of the
coercive field.

So far, the highest thin film quality (smallest combination of low
magnetic damping parameter, low inhomogeneous broadening, and film
thickness below 20~nm) have been reported for thin films grown by the
LPE technique, an extension of the flux method. Garnets have a
non-congruent melting phase and can only be prepared in the form of
single crystals once dissolved in a solvent. The solvent used is
usually a mixture of different oxides elements, mainly PbO and
B$_2$O$_3$, which can eventually enter as impurities in the flux
growth.  The molten mixture is confined in a platinum crucible (inert
with respect to the oxides) placed in an epitaxy furnace above the
saturation temperature, defined as the temperature at which the growth
rate is zero. Subsequently, the GGG substrate with crystallographic
orientation (111) is immersed in the bath. Optimization of the growth
process parameters is achieved by studying the dependencies of the
depositing conditions on the structural, morphological and magnetic
properties. The key to very good growth, is to keep the solution
perfectly homogeneous and the growth rate very slow.  The main problem
is the difficulty of developing a recipe leading to YIG films
homogeneous in both thickness and composition. Indeed, for very thin
layers, the role of the transition layer is essential (chemical
composition) and requires a control of the chemical elements composing
it. Indeed, the influence of this transition layer on the different
contributions to the line width is important. The YIG films, that we
have developed from LPE growth technique, have the following
characteristics: perfect epitaxy (difference of matching parameter
with the substrate is null); spontaneous magnetization almost equal to
that of the bulk ($4 \pi M_s$ of our 19~nm YIG films is about 1.7~kG);
very low magnetic relaxation (damping coefficient less than or equal
than $3.5\times 10^{-4}$); no planar anisotropy and very weak coercive
field ($H_c <3$~Oe); very low roughness (3 \AA$_\text{rms}$).

\begin{table}
  \caption{Summary of the physical properties of the materials used in this study.}
  \begin{ruledtabular}
    \begin{tabular}{c | c c c c }
      YIG & $t_\text{YIG}$ (nm) & $4 \pi M_s$ (G) & {$\alpha_\text{YIG}$} &
      {$\Delta H_0$} (Oe)\\ \hline
      & 19 & $1.67 \times 10^3$ & {$3.2 \times 10^{-4}$}  & {2.5} 
 \end{tabular}
\end{ruledtabular}\label{tab:mat}
\end{table}

\begin{figure}
  \includegraphics[width=8.5cm]{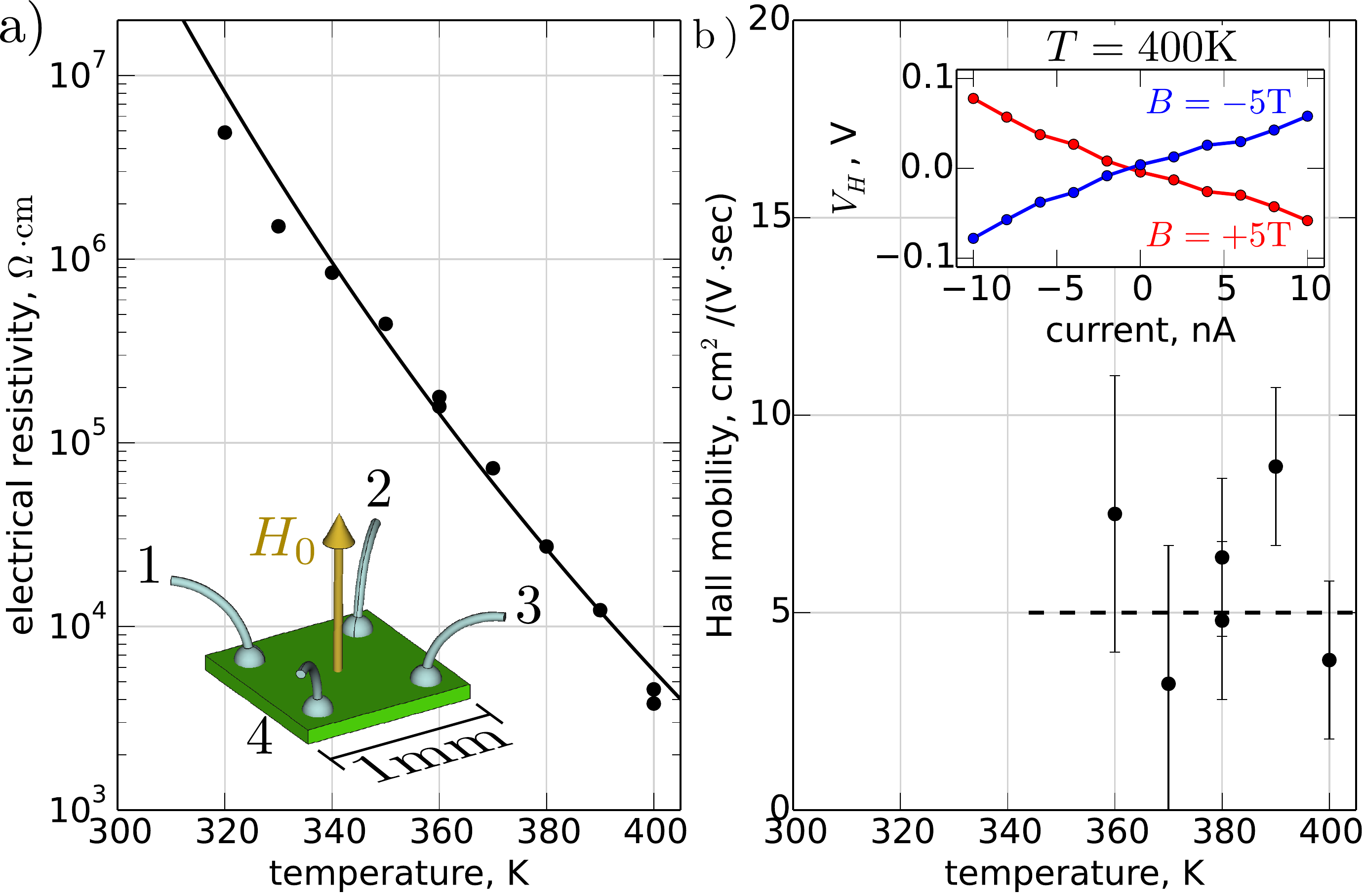}
  \caption{(Color online) Temperature dependence of a) the electrical
    resistivity and b) Hall mobility of 19~nm thick YIG films grown by
    LPE determined by a Van der Pauw four-point probe technique (see
    insert). The solid line in a) is a fit with an activated behavior
    $\exp[ E_g/(2 k_B T) ]$, where $E_g \approx 2$~eV. The insert in
    b) shows the Hall voltage drop $V_{i+1,i+3}$ when the current is
    injected between $I_{i,i+2}$ where $i$ is the contact number
    modulo 4.}
  \label{FIG1}
\end{figure}

In the following we will concentrate on the electronic properties of
LPE grown YIG thin films of thickness $t_\text{YIG}=19$~nm. The
dynamical characteristics of these films are summarized in Table 1.  A
1$\times$1mm$^2$ square slab of YIG is extracted from the batch and
connected along the 4 corners using Al wire bonding. To characterize
the slab we use the van der Pauw four probes method, which is
typically used to measure the sheet resistance of homogeneous
semiconductor films. It allows to eliminates measurement errors
associated with the exact shape of the sample. The four points are
arranged in a clockwise order around the positive field normal shown
schematically inside FIG.\ref{FIG1}a. Because of YIG high impedance,
we have used a Keithley 2636B source-measurement unit in order to draw
very little current (sub-nA range) inside the film. In our analysis,
the GGG substrate will be considered a good insulator (resistivity $>
10^{15} \Omega \cdot \text{cm}$) \cite{Metselaar1978} and its
electrical conductivity will be ignored.

Our measurements are performed at high temperature in the range
[300,400]~K and for different magnetic fields in the range [0,5]~T
applied normally to the sample surface. The temperature range explored
is still well below the Curie temperature of YIG, which is
$T_c=562$~K. We first extract the sheet resistance $R_s$, which
consists in measuring all possible combinations of the
cross-resistance between opposite edges. From the van der Pauw
expression, one can extract $R_s$, whose minimum lays in the couple of
G$\Omega$ range at the highest temperature. From the sheet resistance,
we compute the resistivity $\rho = R_s t_\text{YIG}$ . FIG.\ref{FIG1}
shows the resistivity as a function of temperature. The first
remarkable feature is that the resistivity of YIG at 400K drops to
about $5\times 10^3$~$\Omega \cdot \text{cm}$. Plotting the data on a
semi-logarithmic scale helps to show that the decay of the resistivity
follows an exponential behavior. Fitting a linear slope through the
points on the plot, we infer a band-gap energy of about $E_g
\approx$2~eV, which is about 1~eV lower than the expected band-gap of
pure YIG in bulk form.

Next, we characterize the Hall conductance of our sample. For this, we
now circulate the electrical current along the diagonals $I_{i,i+2}$
and measure the voltage drop along the opposite contacts
$V_{i+1,i+3}$. Here $i$ is the contact number modulo 4, where the
subscript notation is ordered according to the connections to the
high/low binding posts of the current source and voltmeter. The insert
of FIG.\ref{FIG1}b shows the voltage drop measured at 400~K in the
presence of a normal magnetic field of 5~T. To eliminate the
resistivity offset, we have worked out the difference of the voltages
for positive and negative magnetic fields. In our measurement
geometry, the polarity of the Hall voltage is opposite to the magnetic
field direction. It implies that the trajectories of the charge
carriers are deflected in the opposite direction to the current in the
electromagnet, or in other words that the YIG behaves as a $p$-type
conductor. Quantitatively the full variation of the Hall voltage is
about 0.12~V at 10~nA when the field is changed by $H_0=\pm$5~T at
400~K, where the YIG resistivity is $\rho=5\times
10^3$~$\Omega\cdot$cm. This corresponds to a carrier mobility for the
holes of about $\mu_H \approx 5$~cm$^2$/(V$\cdot$sec). We have
repeated the measurement for other temperatures. The measurement at
lower temperature is difficult for 2 reasons. The first one is the
limited voltage range of the sourcemeter, which decreases the upper
current limit that could be injected in the YIG. Another consequence
of the large resistivity, is the associated increase of the time
constant for charging effects. This increases substantially the dwell
time necessary before taking a measurement. Because of these
difficulties, we have limited the measurement range to 40~K below the
maximum temperature. It seems that the temperature dependence of the
mobility as a function of temperature is very small \cite{Bullock1970}
indicating that most of the change in the resistivity comes from a
variation of the electronic density and not of the scattering
time. Such behavior is compatible with what has been found previously
in Ca doped YIG ($p$-type) and this observation is used as a signature
that charge carriers are provided by large polarons
\cite{Larsen1976}. Our study does not conclude if the electrical
conduction occurs in the bulk or if this is a surface effect. This
important question shall be determined in future studies by monitoring
the change in the electrical properties as a function of the YIG
thickness.

\begin{figure}
  \includegraphics[width=8.5cm]{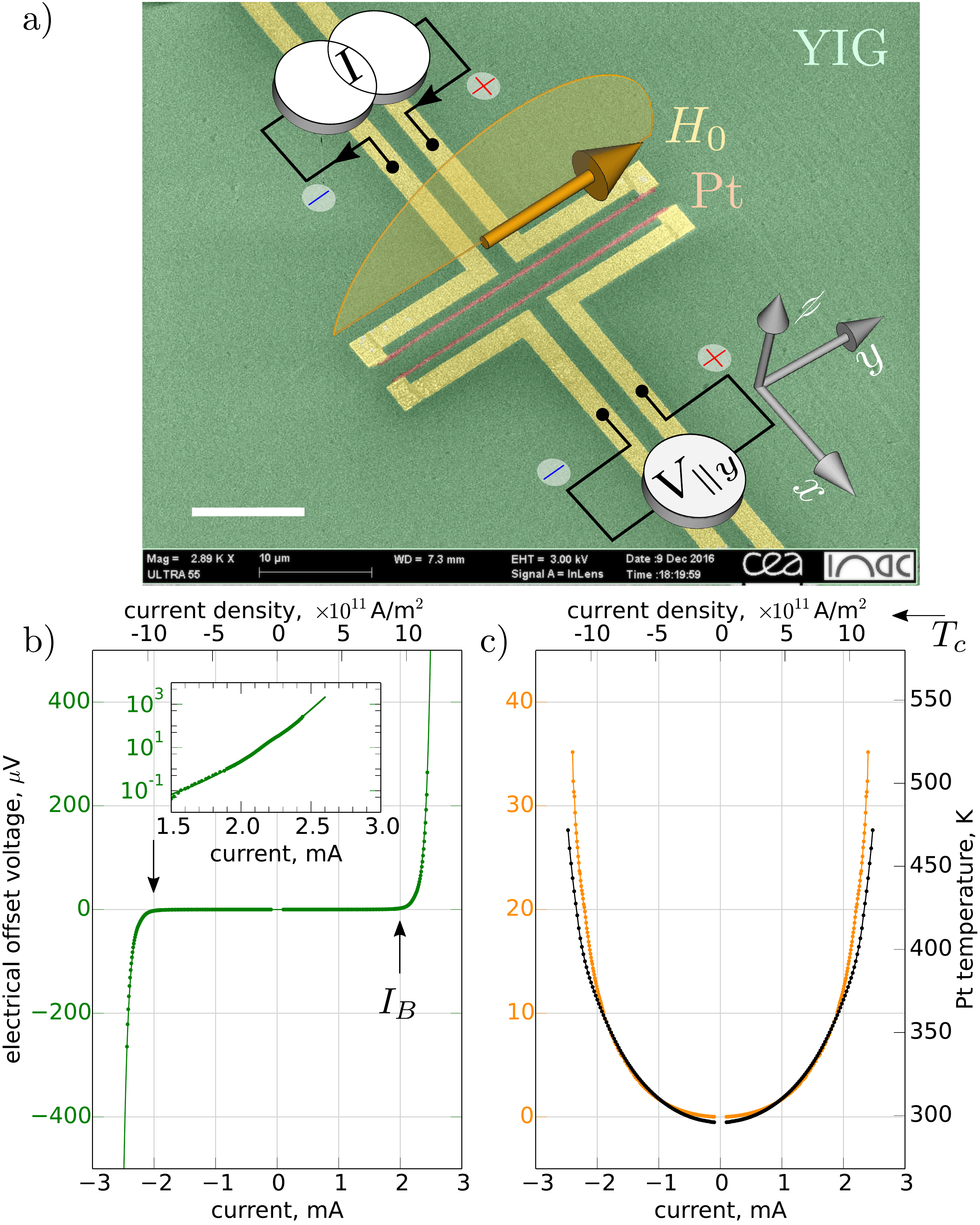}
  \caption{(Color online) Current dependence of the electrical offset
    voltage $V_{\parallel y}$ in a non-local transport device, one
    monitors the voltage along one wire as a current flows through a
    second wire. Panel a) is its microscopy image showing two Pt
    stripes along the $y$-direction (red) (the scale bar is
    10~$\mu$m). The polarity of the current source and voltmeter are
    specified. The YIG magnetization is set along the $y$-direction by
    an external in-plane magnetic field, $H_0=2$~kOe. The offset
    voltage is decomposed in two contributions: b) $(V_{\parallel,+I}
    - V_{\parallel,-I})/2$, odd in current (green), and c)
    $(V_{\parallel,+I} + V_{\parallel,-I})/2$, even in current
    (orange). The solid line in b) is a fit with an exponential
    increase $\exp[-E_g / (2 k_B T)]$, where $E_g \approx 2$~eV. The
    insert is a zoom of the data and fit on a semi-logarithmic
    scale. The black curve in c) shows the increase of relative
    resistance of the Pt used as a temperature sensor. The arrow at
    $i_B=2$~mA indicates the threshold current at which the Ohmic
    losses start to become non-negligible in the spin transport
    experiments.}
  \label{FIG2}
\end{figure}

Next, we investigate the implications of these electrical properties
for the non-local experiments \cite{cornelissen16}, where one monitors
the transport properties between two parallel metal wires deposited on
top of YIG. More precisely, one measures the voltage along one wire
(the detector) as a current flows through a second wire (the
injector). FIG.\ref{FIG2}a shows a microscopy image of the electrode
pattern on top of the YIG. In these lateral devices, the two parallel
wires are made of Pt (see two red lines in FIG.\ref{FIG2}a along the
$y$-direction) and the same devices have also been used to investigate
the spin conduction properties of YIG. For the lateral device series
used herein, the Pt wires are 7~nm thick, 300~nm wide, and 30~$\mu$m
long. Since different Pt wires (thickness and length) have been
deposited between different samples, comparison of the results should
be done by juxtaposing data obtained with identical current densities
(provided in the upper scale).  The total resistance of the Pt wire at
room temperature is $R_0=3.9$~k$\Omega$, corresponding to a Pt
resistivity of 27.3 $\mu \Omega$.cm. Although the analysis below
concentrates on a particular lateral device, whose Pt wires are
separated by a gap of 0.4~$\mu$m, these measurements have been also
performed on a multitude of other devices patterned on two different
LPE YIG film batch of similar thickness. In the following, we shall
explicitly clarify the effects, that are generic to the YIG films. In
our measurement setup the current is injected in the device only
during 10~ms pulses using a 10\% duty cycle. This pulse method is very
important in order to limit heating of the YIG and substrate. The
increase of resistance $R_I$ of the Pt wire is monitored during the
pulse. The result is shown in FIG.\ref{FIG2}c (right axis), where we
have plotted $\kappa_\text{Pt} (R_I-R_0)/R_0$ as a function of the
current $I$, with the coefficient $\kappa_\text{Pt} = 254$~K specific
to Pt \cite{landolt70}. The result is shown in FIG.\ref{FIG2}c using
black dots. For information purposes, we have also marked on the plot
the position of the Curie temperature $T_c$. In the following, we
shall assume that the local YIG temperature is identical to that of
the Pt (\textit{i.e.}  assuming negligible Kapitza resistance
\footnote{A more precise analysis should consider the possibility that
  the temperature differs between the YIG and the Pt.}). One can use
this plot to estimate the temperature effects on the electrical
properties. At $I=2$~mA, which corresponds to current density of about
$10^{12}$A$\cdot$m$^2$ circulating in the Pt injector wire, the
temperature of the YIG has increased to about 370~K during the
pulse. At this temperature, the YIG resistivity drops into the
sub-$10^5$~$\Omega \cdot \text{cm}$ range according to
FIG.\ref{FIG1}a, which corresponds to a sheet resistance of about
50~G$\Omega$. Considering now the lateral aspect ratio of the device,
this amounts to an electrical resistance of YIG of the order of the
G$\Omega$ between the two wires. The leakage current inside the Pt
detector wire, whose impedance is about 6 orders of magnitude smaller
than the one of YIG, starts thus to reach the sub-nA range, which is
comparable to the induced currents produced by inverse spin Hall
effects.

To resolve this effect in our lateral device, we propose to measure
the non-local voltage with the magnetization set precisely parallel to
the Pt wire. This configuration switches off completely any
sensitivity to spin conduction. To align the magnetization with the
wire, an external in-plane magnetic field of $2$~kOe is applied along
the $y$-direction as shown in FIG.\ref{FIG2}. The induced offset
voltage is decomposed in two contributions: b) one $(V_{\parallel,+I}
- V_{\parallel,-I})/2$, which is odd in current (green) and the other
c) $(V_{\parallel,+I} + V_{\parallel,-I})/2$, which is even in current
(orange).

We first concentrate on the odd contribution of the offset shown in
green in FIG.\ref{FIG2}b. For the odd offset it is always observed
that, within our convention of biasing the high/low binding posts of
the current source and voltmeter in the same direction
(cf. \encircle{$+$} and \encircle{$-$} polarities in FIG.\ref{FIG2}a),
the sign of $(V_{\parallel,+I} - V_{\parallel,-I})/2$ is positive for
positive current and negative for negative current. It implies that
$(V_{\parallel,+I} - V_{\parallel,-I})\cdot I > 0$, which means that
the voltage drop is produced by dissipation. One should emphasize that
this sign is opposite to the voltage produced by the inverse spin Hall
effect (cf. FIG.2e in ref.\cite{Thiery2017}). On the figure scale, we
observe that the odd offset increases abruptly above $I_B=2$~mA
(corresponding to a YIG temperature of 370~K). This offset actually
follows an exponential growth as shown in the insert using a
semi-logarithmic scale. The solid line in FIG.\ref{FIG2}b is a fit
with an exponential increase $\exp[-E_g/(2 k_B T)]$, where the
temperature $T$ is extracted from the Pt resistance (cf. black dots in
FIG.\ref{FIG2}c). From the fit, we extract the local band-gap $E_g
\approx 2$~eV, which is the same as that extracted from the
resistivity. We then evaluate quantitatively the expected signal from
current leakage through the YIG. At $I=2.2$~mA, when the temperature
of the YIG reaches $T\approx 390$~K, the YIG resistivity drops to
about $10^4$~$\Omega \cdot \text{cm}$. Considering the equivalent
circuit, this will produce a difference of potential of 50$\mu$V on
the detector circuit, which is consistent with the observed
signal. This confirms that the odd offset voltage in our non-local
device is purely produced by the decrease of the YIG electrical
resistivity. Note that this offset voltage drops very quickly with
decreasing current pulse amplitude. As shown in the insert of
FIG.\ref{FIG2}b, it decreases by an order of magnitude, when
$I=2.0$~mA (corresponding to a YIG temperature of $T\approx
370$~K). At this level, the offset starts to become of the same order
of magnitude as the spin signal in these devices. We have thus
indicated by an arrow in FIG.\ref{FIG2}b, $i_B=2$~mA (\textit{i.e.}
current densities of approximately $10^{12}$~A$\cdot$m$^2$), the
threshold current at which the electrical leakage starts to become
important in the spin transport experiments.

We then move to the even contribution of the offset shown in orange in
FIG.\ref{FIG2}c. One observes that this contribution always follows
the Joule heating of the Pt wire, so it is linked to the induced
thermal gradient. It is ascribed to thermoelectric effects produced by
a small temperature difference at the two Pt$|$Al contacts of the
detector circuit. This difference is produced by small resistance
asymmetries along the Pt wire length, which induce one end to heat
more than the other end. Considering the Seebeck coefficient of
Pt$|$Al, of 3.5$\mu$V/K \cite{landolt70}, the offset measured at
$I=2$~mA, corresponds to a temperature difference of less than
3~$^\circ$C between the top and bottom contact electrodes, while the
wire is being heated by almost 70~$^\circ$C. These asymmetries in the
temperature difference are expected to vary from one device to the
other and this is precisely what is observed: the ratio between the
even contribution of the offset and the temperature increase of the Pt
wire fluctuates and even changes sign randomly between different
lateral devices.

\begin{figure}
  \includegraphics[width=8.5cm]{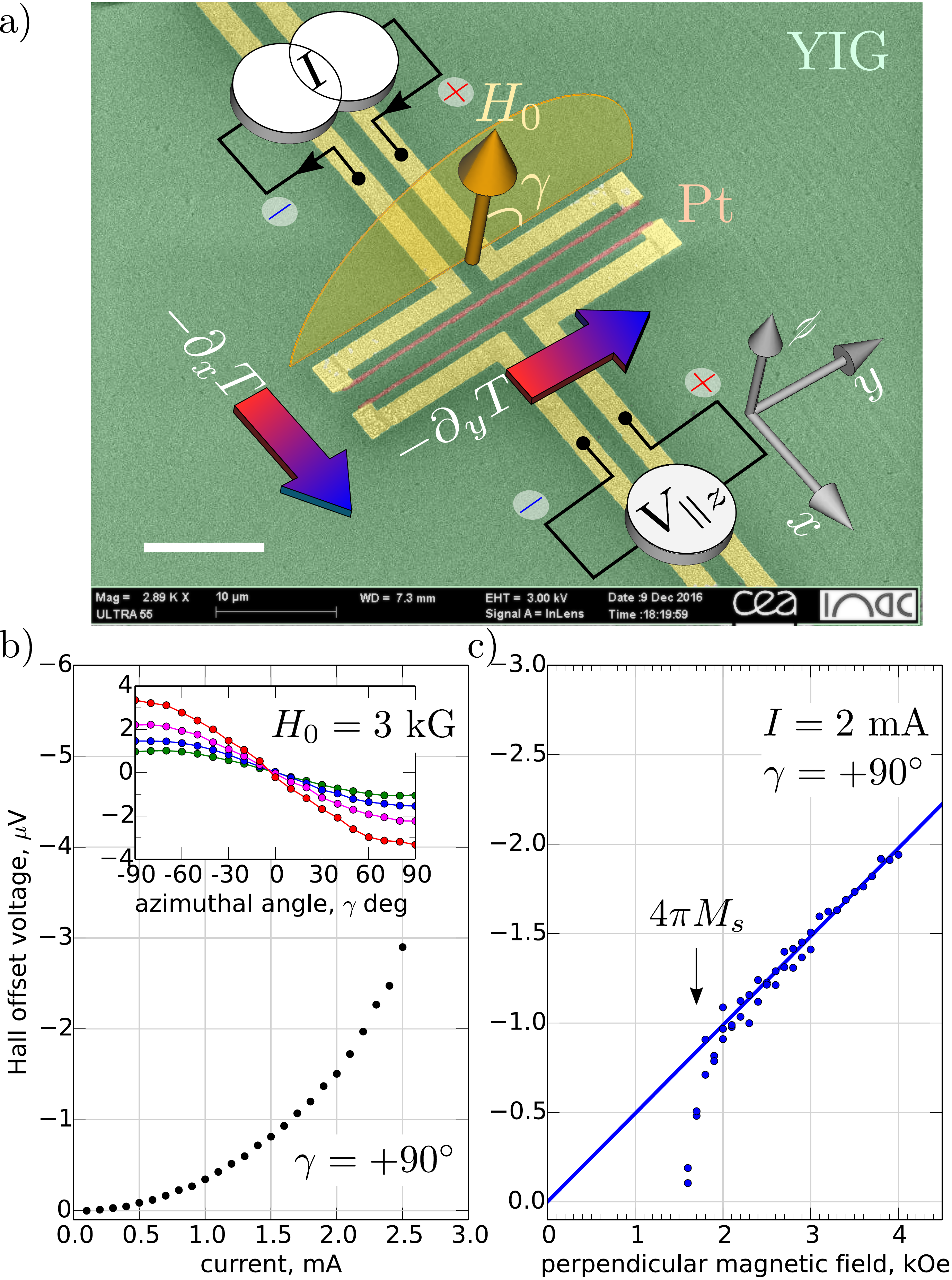}
  \caption{(Color online) a) Schematic of the the Righi-Leduc effect
    produced on a $p$-type conductor magnetized out-of-plane. The
    large in-plane temperature gradient $\partial_x T$ produced by
    Joule heating creates a temperature gradient $\partial_y T$ along
    the wire when the sample is subject to an out-of-plane magnetic
    field. b) Current and c) magnetic field dependence of the Hall
    offset voltage $V_{\parallel z}$ produced in the non-local
    transport device. Panel a) shows the variation of the
    $\Delta$-signal as function of $I$ when $H_0=\pm 3$~kG. The inset
    of b) shows the out-of-plane angular variation for different
    current between [1.9,2.5]~mA (step 0.2~mA). Panel b) shows the
    magnetic field dependence of the $\Delta$-signal measured when the
    field is oriented along $+\hat z$  ($\gamma=+90^\circ$).  }
  \label{FIG3}
\end{figure}

Although the offset voltage is independent of the external magnetic
field direction when the latter is rotated in-plane, it is in
principle sensitive to the out-of-plane component through the Hall
effect. This transverse transport in YIG is attracting a lot of
interest and several recent papers address the issue of transverse
magnon transport effects in magnetic materials, such as the magnon
Hall effect \cite{Onose2010} or the magnon planar Hall effect
\cite{Liu2017}. The sensitivity to magnons effects can be eliminated
from the measurements by keeping the magnetization vector in the
$yz$-plane containing the interface normal and the length of the Pt
stripe and thus this configuration selects only the transverse
transport properties carried by the electrical charge. To extract the
Hall voltage, we shall only consider the $\Delta$-signal, defined as
the difference between the measured voltage for two opposite
polarities of the external magnetic field $\pm H_0$, $\Delta =
(V^+_\parallel-V^-_\parallel)/2$. This signal would be also sensitive
to the spin Seebeck effects if the YIG magnetization had a
non-vanishing projection along the $x$-axis \cite{Thiery2017}.  The
insert of FIG.\ref{FIG3}b shows the angular dependence as a function
of the azimuthal angle, $\gamma$, being defined in FIG.\ref{FIG3}a.
On all the devices, we observe that the $\gamma$-dependence of the
$\Delta$-signal is: maximum when the field is applied along the
$z$-direction; odd in magnetic field; and negative when $\gamma = +
\pi/2$. Moreover, the $\Delta$-signal increases with both increasing
current density and increasing external magnetic field. It is worth to
note at this point that the offset voltage produced by the
perpendicular magnetic field is about two orders of magnitude smaller
than that of the in-plane direction. The observed quadratic dependence
with $I$ in FIG.\ref{FIG3}b suggests that this signal should be
associated to Joule heating and thus to particle flux induced by
thermal gradients. The observed linear dependence with $H_0$ in
FIG.\ref{FIG3}c suggests that this signal should be associated to a
normal Hall effect and not to an anomalous Hall effect linked to
$M_s$. Indeed a fit of the high field data leads to a straight line
that intercepts the origin, while anomalous Hall effect would have
lead to a finite intercept. One also observes in FIG.\ref{FIG3}c a
departure from this linear behavior below the saturation field. This
is because, below saturation, a component of the magnetization could
point in the $x$-direction, hereby switching on the sensitivity to the
spin Seebeck effect, which is a stronger positive signal in these
devices.

Next we discuss more in details the potential origin of the
$\Delta$-signal. First, as noted in the previous paragraph, the source
is the incoming flux of charge carriers produced by a thermal
gradient. This gradient is in the $x$-direction, through the potent
Joule heating of the injector. There is in principle an electrical
voltage produced in the $y$-direction associated with this incoming
flux through the electrical Nernst effect. Our device geometry
effectively shunts both contacts with a relatively low impedance Pt
wire, acting as a voltage divider, which reduces drastically any
sensitivity to the Nernst effect. One should mention at this point the
recently reported spin Nernst effect \cite{Meyer2017}. But this signal
should be maximum when the magnetization is parallel to the
$y$-direction, while the signal that we discuss here is maximum when
the magnetization is parallel to the $z$-direction. We propose here a
different scenario to explain our data. Since our measurement of the
even offset in FIG.\ref{FIG2}c seems to indicate that the two
thermocouples provided by the Pt$|$Al contacts at both ends of the
detector Pt wire are sensitive to temperature difference along the
$y$-direction, the Hall offset signal measured in FIG.\ref{FIG3} can
thus be due to a thermal gradient in the $y$-direction (Righi-Leduc
effects \cite{Madon2016}). Although a definite quantitative proof
would require some additional measurements, in the following we shall
check that this explanation is consistent with the data.


Firstly, this explanation is consistent with the $I$ and $H_0$
behavior. Secondly, it also has the correct sign. Since the
$\Delta$-signal is negative for $\gamma=+\pi/2$, this implies that
$\partial_x T$ and $\partial_y T$ have the same sign when the field is
positive. This is the signature of a $p$-type doping in agreement with
FIG.\ref{FIG1}b. Concerning the amplitude of these thermal gradients,
at $I_B=2$~mA, we evaluate the temperature of the YIG underneath the
injector and the detector by measuring the value of the Kittel
frequency at these two positions using $\mu$-BLS
spectroscopy. Comparing these values between FIG.4a and FIG.4b of
ref.\cite{Thiery2017} allows to estimate the decrease of spontaneous
magnetization under the injector (-290~G) and the detector
(-110~G). If one uses the fact that YIG magnetization decreases by
4G/$^\circ$C in this temperature range, we find that at $I=2$~mA the
temperature of the YIG underneath the injector has increased by
+73$^\circ$C, while the temperature of the YIG underneath the detector
has increased by +27$^\circ$C. The fact that these values agree with
the increase of temperature of the Pt resistance confirms that the
Kapitza resistance is probably small. Using the gap of 0.4$\mu$m
between the 2 Pt wires as the area where the thermal gradient along
$x$ occurs, we find that $\partial_x T = -130^\circ$C/$\mu$m. Using a
value of $\mu_H=+5$~cm$^2$/(V$\cdot$sec) for the Hall mobility, this
would produce a transverse gradient of $\partial_y T = \mu_H
H_0 \partial_x T = -2\times 10^{-2}$~$^\circ$C/$\mu$m in a
perpendicular magnetic field of 3~kG. Recalling that the length of the
Pt wires is 30$\mu$m long, this would produce a voltage of 2$\mu$V,
using the thermoelectric coefficient of Pt$|$Al. So in the end, the
expected signal amplitude is of the same order of magnitude as the
Hall offset measured at $I_B=2$~mA.

In summary, we have shown that high quality YIG thin films grown by
LPE behave as a large gap semiconductor at high temperature due to
the presence of a small amount of impurities inside the YIG.  In our
case, we observe that the resistivity drops to about $5\times
10^3$~$\Omega \cdot \text{cm}$ at $T=400$~K. These results are
important for non-local transport exploring the spin conductivity,
especially in cases where the YIG is subject to a large amount of
defects like in amorphous materials, or when improper cooling of the
sintered product leads to the additional formation of Ohmic grain
boundaries. In non-local transport devices, these electrical
properties are responsible for the abrupt emergence of an odd offset
voltage at large current densities as well as a temperature gradient
along the wire proportional to the perpendicular component of the
magnetic field. These results emphasize the importance of reducing
drastically the Joule heating by using a pulse method, when
investigating spin transport in YIG in the strong out-of-equilibrium
regime. For our devices, these electrical properties start to become
non-negligible for spin transport studies, when the YIG temperature is
heated $>370$~K, which corresponds in our case to injecting a current
density $>1.0\times 10^{12}$~A$\cdot$m$^2$ in the Pt (or $I>2$~mA for
these samples). We add that non-local voltage produced by Ohmic losses
in the YIG are easily separated from the non-local voltage produced by
spin transport. Firstly, the two signals have opposite
polarities. Secondly, only the latter varies with the orientation of
the magnetization in-plane, as first demonstrated by the Groeningen
group \cite{cornelissen16}.

\begin{acknowledgments}
  This research was supported by the priority program SPP1538 Spin Caloric
  Transport (SpinCaT) of the DFG and by the program Megagrant
  14.Z50.31.0025 of the Russian ministry of Education and Science. VVN
  acknowledges fellowship from the emergence strategic program of UGA,
  and Russian program of competitive growth of KFU.

\end{acknowledgments}

%

\end{document}